\documentclass[prl,twocolumn]{revtex4}
\usepackage{graphicx}
\renewcommand{\v}[1]{{\bf #1}}

\newcommand{\w}{{\omega}}

\def\eqa{\begin{eqnarray}}
\def\eea{\end{eqnarray}}
\newcommand{\eq}{\begin{equation}}
\newcommand{\ee}{\end{equation}}

\newcommand{\<}{\langle}
\renewcommand{\>}{\rangle}

\renewcommand{\Im}{{\rm Im}}

\newcommand{\ua}{\uparrow}
\newcommand{\da}{\downarrow}
\newcommand{\ra}{\rightarrow}

\newcommand{\del}{\delta}
\newcommand{\Del}{\Delta}

\begin{document}

\title{Theory of high energy features in angle-resolved photo-emission spectra of hole-doped cuprates}
\author{Qiang-Hua Wang, Fei Tan and Yuan Wan}
\address{National Laboratory of Solid State Microstructures \& Department of Physics, Nanjing University,
Nanjing 210093, China}


\begin{abstract}
The recent angle-resolved photoemission measurements performed up
to binding energies of the order of 1eV reveals a very robust
feature: the nodal quasi-particle dispersion breaks up around
0.3-0.4eV and reappears around 0.6-0.8eV. The intensity map in the
energy-momentum space shows a waterfall like feature between these
two energy scales. We argue and numerically demonstrate that these
experimental features follow naturally from the strong correlation
effects built in the familiar t-J model, and reflect the
connection between the fermi level and the lower Hubbard band. The
results were obtained by a mean field theory that effectively
projects electrons by quantum interference between two bands of
fermions instead of binding slave particles.
\end{abstract}

 \maketitle

Recently, several groups performed independent measurements of the
electron structure in hole-doped cuprates at binding energies $E$
up to $1eV$.\cite{feng,lanzara,ding,zhou,valla} They observed that
starting from the nodal Fermi point the nodal-direction
quasi-particle dispersion breaks up near the momentum
($\pi/4$,$\pi/4$) while approaching the zone center at $E>E_1\sim
0.3-0.4$ eV. The dispersion curve then drops in a
waterfall-fashion up to $E>E_2\sim 0.6-0.8$ eV, where spectral
weights reappear while dispersing toward the zone center. The
waterfall also appears in the antinodal direction near
($\pi/2$,0). These features are observed in the under-, optimal as
well as over-doped regimes, below or above the superconducting
transition temperature in hole-doped cuprates. In contrast, this
feature does not appear in manganites,\cite{zhou} signifying the
unique property of cuprates. Given the robust phenomenology, the
mechanism should be independent of pairing. Phonons were seen to
play important roles at low energy scales\cite{shen} and at low
doping levels\cite{polaronshen}. However, it is not clear whether
they could cause a dynamical gap of the order of electron volt. It
is also not clear whether the polaron physics\cite{nagaosa}
applies where doped holes are already very metallic. The purpose
of this paper is to show that the robust waterfall feature may
reflect the generic property of one-band t-J model, serving as a
connection between the low energy quasi-particles and the residual
lower Hubbard band (LHB) at higher binding energies dominated by
local Mottness. In reaching this conclusion, we deal with the
strong correlation effect built in the t-J model by projecting
electrons with quantum interference between two bands of fermions.
We argue that this procedure satisfies the local sum rules for
projected electrons already at the mean field level, and is
therefore able to pick out the higher binding energy degrees of
freedom.

\begin{figure}
\includegraphics[width=8.5cm]{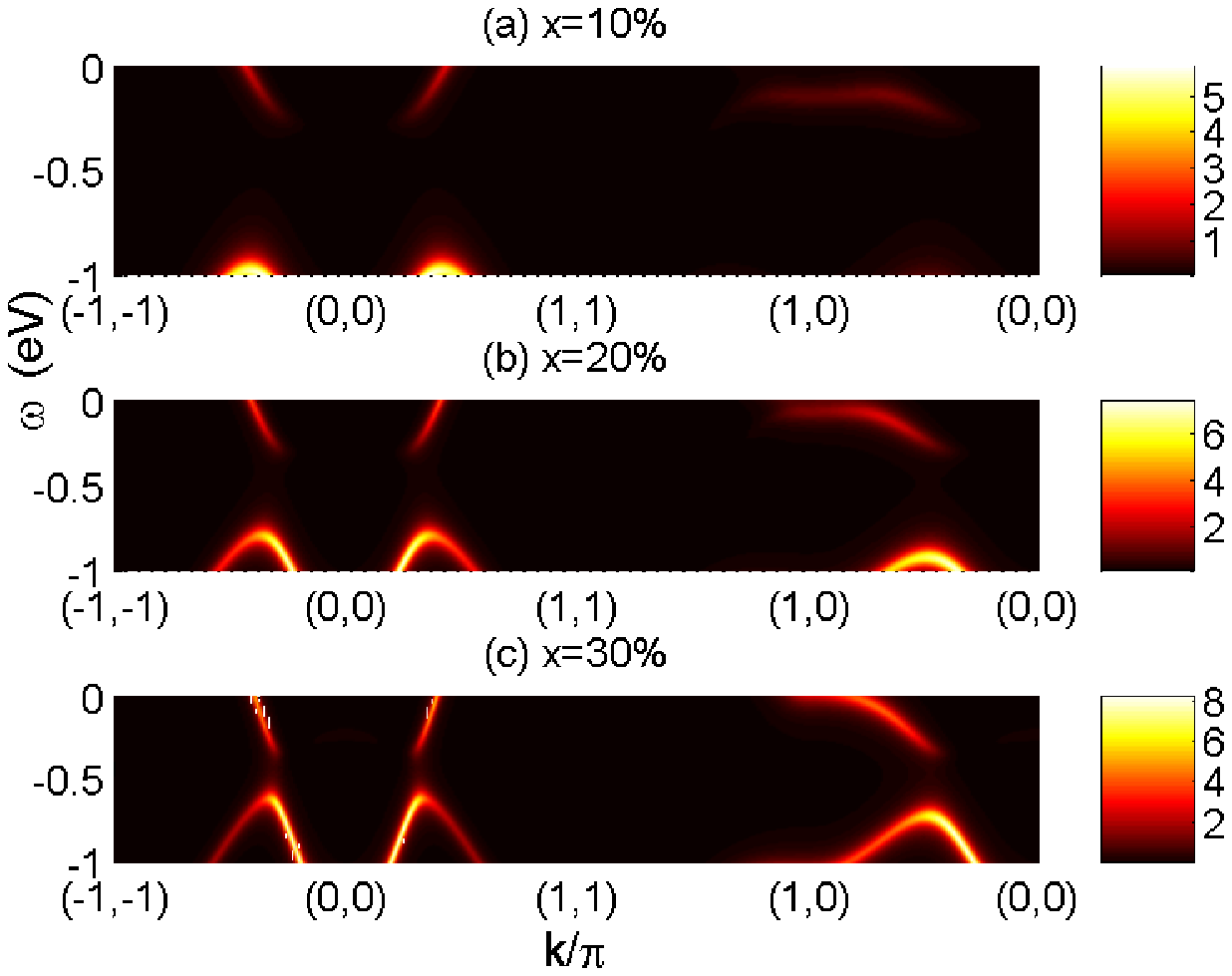}
\caption{(Color online) Intensity maps of $A(\v k,\w)$ in the
energy-momentum space (along high symmetry cuts). The doping level
is (a) $x=10\%$, (b) $x=20\%$, (c) $x=30\%$. The intensity scales
with the hotness of the color, as shown by the color bars. See the
text for details.}
\end{figure}

\begin{figure}
\includegraphics[width=8.5cm]{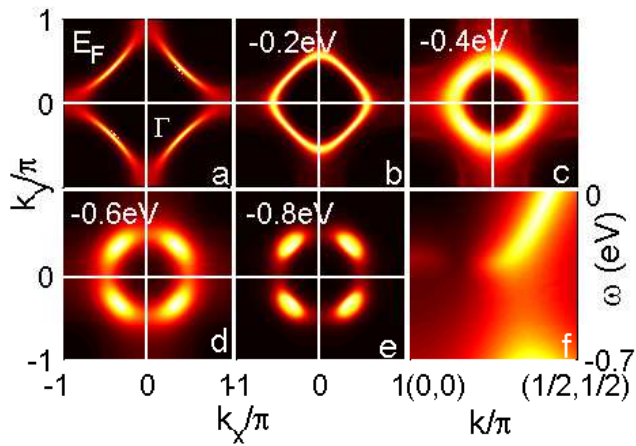}
\caption{(Color online) Panels (a)-(e): Intensity maps of $A(\v
k,\w)$ in the momentum space for $x=20\%$. The binding energies
increases linearly from 0 to 0.8eV. Panel (f): A view field
extracted from Fig.1(b) but re-plotted in a log scale for the
intensity to reveal the waterfall.}
\end{figure}

We first summarize the main results in comparison with the
angle-resolved photoemission (ARPES) data. Using parameters
suitable for hole-doped cuprates, we calculated the electronic
spectral weight $A(\v k,\w)$ as a function of momentum $\v k$ and
binding energy $E=-\w$. Along the cut $(-\pi,-\pi)\ra (0,0)\ra
(\pi,\pi)\ra (\pi,0)\ra (0,0)$ this is presented in Figs.1 as
color intensity plots at doping levels $x=10\%$ (a), $20\%$ (b)
and $30\%$ (c). At low binding energies, the nodal quasi-particle
spectral weight fades away while approaching the zone center,
producing a break up roughly at $\pm (\pi/4,\pi/4)$. On the other
hand, in the antinodal direction, there is also a break up of the
dispersion near $(\pi/2,0)$. The energy scale of the this break is
around $E_1=0.3eV$. The low energy spectral weight scales with the
doping level, as seen from the increasing brightness with
increasing doping in Figs.1. This is the general behavior of doped
Mott insulators. At higher binding energies $E>E_2\sim 0.6-0.8eV$,
spectral weight reappears and behaves in disguise as the missing
low energy dispersion pushed to higher binding energies. We note
that in Figs.1 the spectral weight in the waterfall energy window
is weak but nonzero.\cite{note} In order to reveal the waterfall
in our case we plot $A(\v k,\w)$ as intensity maps for $x=20\%$ in
the momentum space at different energies in Figs.2(a-e). We also
extract a view field from Fig.1(b) and re-plot it in Fig.2(f)
using a log scale for the intensity. The waterfall feature and the
pinning of the waterfall spectral weight in the momentum space are
now obvious. In ref.\cite{lanzara} the pinned momentum boundary
was emphasized as along a linearly-halved magnetic zone boundary.
However, this does not seem to be a universal feature in all
cuprates, even though the waterfall indeed appears near this
boundary.\cite{zhou,feng,ding,valla} The breaking momentum is
doping dependent as explicitly pointed out in ref.\cite{valla}.
This is also the case in Figs.1. We conclude that the qualitative
features of our results are in remarkable agreement with the data.

We add that the waterfall energy window decreases before it
disappears at extremely high doping levels. The tendency is seen
in Figs.1. In the opposite limit, as $x\ra 0$ the high energy
features remain (not shown here) and this is why we believe that
they reflects the LHB. In the experimental case,\cite{undoped} the
waterfall seems to also exist in this limit, together with the
nodal dispersion for binding energies larger than but close to
$0.35eV$. The latter may be identified as the low energy
dispersion in our case but folded by the anti-ferromagnetic order,
which will be studied elsewhere.

As for the mechanism of the waterfall, in our case it is merely a
manifestation of a connection between the Fermi level and the
parent LHB. The higher-energy dispersion is the residual of
un-doped systems. By doping a Mott insulator, the Fermi level does
not have to sink immediately within the LHB (the upper Hubbard
band is pushed to infinity in the t-J model), but remains above
the parent band. Spectral weights are transferred from the parent
band to the Fermi level. The amount of spectral transfer increases
with doping, and is highly anisotropic in the momentum space due
to the dispersion of the parent lower band. In the theory to be
described, an effective break up in the low binding energy
dispersion occurs due to the destructive quantum interference
effect near the zone center, while the spectral weight in the
waterfall energy window arises from the tails of excitations near
the window boundaries where van Hove singularities appear in the
Bogoliubov de Gennes bands. One of the singularities is already
visible around the tips of the high energy dispersion in Figs.1.

The theoretical starting point is the familiar one-band t-J model,
\eqa H=-\sum_{i,n,\sigma} t_{n}(c_{i\sigma}^\dagger
c_{i+n,\sigma}+{\rm h.c.})+J\sum_{\<ij\>} S_i\cdot S_j,\label{HtJ}
\eea where no-double-occupancy is implicitly assumed. Here
$n=\hat{x},\hat{y},\hat{x}+\hat{y},\hat{x}-\hat{y},2\hat{x},2\hat{y}$
denotes relevant independent bonds of hopping. This model has
received tremendous efforts since the discovery of cuprate
superconductors. One approach to deal with the projected electrons
is to regard the electron as a composite particle of a boson and a
fermion.\cite{slave} The slave-boson mean field theory (SBMFT)
yields a fermion band re-normalized by the doping level. The
dispersion of quasi-particles in such a theory does not break up.
One could go beyond the mean field level by integrating out
internal gauge fields coupling to the holons and spinons, but no
controllable approximation is available for this purpose. On the
other hand, a similar re-normalized mean field theory (RMFT) is
obtained by directly resorting to Gutzwiller projecting a trial
BCS-like wave function.\cite{rmft} A common feature of SBMFT and
RMFT is the conventional idea of filling the one-band fermion
levels. This restricts the extent of many-particle entanglement,
and predicts that the Fermi level sinks immediately within the LHB
once a Mott insulator is doped. We add that there are also efforts
to bosonize the t-J model,\cite{bosonization} but digging out the
fermionic excitations turns out to be difficult, except possibly
for the one-hole problem.\cite{onehole}

Recently, a new representation of the t-J model in terms of
two-band fermions was proposed.\cite{wen} The idea is to let one
of the band, say the $p$-fermion band, to carry the spin and
charge of doped particles, while the other singly-occupied
half-filled band, say the $f$-fermion band, to reflect the neutral
spin background. By enforcing the condition that each $p$-fermion
pairs up with an $f$-fermion into an on-site spin-singlet the
allowed Hilbert space can be mapped exactly to that of the t-J
model. In the context of hole-doped cuprates, the $p$-band carries
oxygen holes (in the non-bonding band) while the $f$-band carries
copper holes, and the $p$-$f$ spin-singlet is nothing but the
well-known Zhang-Rice singlet,\cite{zrs} which maps to the vacancy
in the effective one-band t-J model. To cope with this analogy, we
switch from now on to the hole picture.

According to the two-band fermion representation, a physical hole
removal (or equivalently an electron creation) is such that we
annihilate a singlet $p$-$f$ pair, and insert back a $f$-fermion
with the right spin quantum number, \eqa c_\sigma^\dagger=d_\sigma
=\frac{1}{\sqrt{2}}\sum_{\sigma'}\epsilon_{\sigma\sigma'}
f_{\sigma'}^\dagger (f_\da p_\ua-f_\ua p_\da),\eea where
$\epsilon$ is a $2\times 2$ antisymmetric tensor. No double
occupation of $f$-fermions is implicitly assumed. We might also
need to require no double occupancy of the $p$-fermions. But this
can be relaxed for two reasons. First, the density of $p$-fermions
is low so that the probability of their double occupancy is small.
Second, the $p$-fermions eventually tries to form singlet pairs
with the $f$-fermions, which is automatically optimized by no
double occupancy of $p$-fermions. The hamiltonian can therefore be
rewritten as, \eqa H=\sum_{\<i,n\>\sigma}t_n (d_{i\sigma}^\dagger
d_{i+n,\sigma}+{\rm h.c.})+J\sum_{\<ij\>} S^f_i\cdot S^f_j
E^p_iE^p_j.\eea Here $S^{f}$ denotes the spin carried by
$f$-fermions, and $E^p=p_\ua p_\ua^\dagger p_\da p_\da^\dagger$
picks up the site with no $p$-fermions at all. Clearly the
hamiltonian conserves the $f$-fermion occupancy on each
site.\cite{su2} As usual, the constrain for $f$-fermions is
relaxed to a global lagrangian multiplier at the mean field level.
The number of $p$-fermions is only globally conserved, and can be
fixed by a chemical potential. The total number of $p$-$f$
spin-singlet pairs is also conserved by the hamiltonian, and can
therefore be fixed by a Lagrangian multiplier.

We now consider the mean field decoupling of the kinetic term. In
the spirit of infinite dimension, in the mean field average
$\<d_{i\sigma}^\dagger d_{j\sigma}\>_0$ we only retain those
contributions with only one inter-site Wick contraction. This is
equivalent to Wick contract within the $d$-operator, leaving
single $f$- or $p$-operator. Starting from the identity \eqa
d_\sigma =&
&\frac{1}{\sqrt{2}}\sum_{\sigma'}\epsilon_{\sigma\sigma'}
f_{\sigma'}^\dagger (f_\da p_\ua-f_\ua p_\da)\nonumber\\ =&
&\frac{1}{\sqrt{2}}( p_\sigma\sum_{\sigma'} f_{\sigma'}^\dagger
f_{\sigma'}+ \sum_{\sigma'} f_{\sigma'}^\dagger
p_{\sigma'}f_\sigma),\label{projectivec}\eea we realize that at
the mean field level $d_\sigma$ could be expressed as a linear
superposition, \eqa d_\sigma \sim \frac{1}{\sqrt{2}} (p_\sigma+
\phi
f_\sigma+\del\sum_{\sigma'}\epsilon_{\sigma\sigma'}f^\dagger_{\sigma'}).
\label{mfd}\eea Here $\phi=\<\sum_\sigma f_\sigma^\dagger
p_\sigma\>_0$ is the mean field p-f particle-hole amplitude, and
$\del=\<p_\da f_\ua -p_\ua f_\da\>_0$ is the mean field p-f
pairing amplitude. Indeed $\phi f_\sigma$ and $\del
\sum_{\sigma'}\epsilon_{\sigma\sigma'}f_{\sigma'}^\dagger$ carry
the same spin and charge quantum numbers as $p_\sigma$ does, given
the fact that $f_\sigma$ are charge neutral spinon degrees of
freedom.

We observe that the constrain on the number of $p$-$f$ singlet
pairs is equivalent to $\<S^f\cdot S^p\>=-3x/4$, where $S^p$ is
the $p$-fermion spin, if the average were performed exactly by
taking into the no-double occupancy of $f$-fermions. In terms of
the mean field average we have $\<S^f\cdot S^p\>=2\<S^f\cdot
S^p\>_0$ where the re-normalization factor of $2$ accounts for the
effect of projection on $f$-fermions. In a mean field decoupling
(with no static spin moments), $\<S^f\cdot
S^p\>_0=-3(\phi^2+\del^2)/8$. We therefore arrive at an important
constrain on $\phi$ and $\del$ as,\eqa
|\phi|^2+|\del|^2=x.\label{constrain}\eea This condition must be
embedded in the mean field theory. As one of the important
differences to our case, it is relaxed in the numerical
calculations in ref.\cite{wen}, where $\del$ is set to zero, and
$\phi=0$ is taken as the signature of the spin-charge separated
pseudo-gap phase. The latter phase is absent due to the constrain
in our case.

From now on we refer $d$-operators in the mean field sense as in
Eq.(\ref{mfd}). We argue that combining with the above constrain
they respect the average but exact local sum rules of doped Mott
insulators, and as such {\em they work as the quasi-particle
operator in doped Mott insulators}. We first recall that in the
nonmagnetic uniform states of the t-J model, the electron occupied
weight is $1-x$, the electron unoccupied weight is $2x$, and they
sum up to yield a total weight $1+x$.\cite{sumrule} These follows
from simple considerations of the projected electron operators
$c_\sigma c_{\bar{\sigma}}c_{\bar{\sigma}}^\dagger$ and
$c_\sigma^\dagger c_{\bar{\sigma}}c_{\bar{\sigma}}^\dagger$. In
our case, the total spectral weight of $d$'s is given by \eqa
\sum_\sigma\<\{d_\sigma,
d^\dagger_\sigma\}\>_0=1+|\phi|^2+|\del|^2=1+x.\eea The hole
unoccupied weight, or in the reversed picture, the electron
occupied weight, is given by \eqa \sum_\sigma\<d_\sigma
d_\sigma^\dagger\>_0=1-(x+|\phi|^2+|\del|^2)/2=1-x.\eea The
electron unoccupied weight is just the difference of the above two
quantities, \eqa \<d_\sigma^\dagger
d_\sigma\>_0=(x+3|\phi|^2+3|\del|^2)/2=2x.\eea {\it These sum
rules guarantee that we can use the mean field Greens function of
$d$-operators even for local operators and therefore capture high
energy features}, with the projection arising from quantum
interference between two-bands of fermions. In contrast, the SBMFT
and RMFT calculate the electron greens function as
$G_c=G_{coh}+G_{inc}$ with an unknown incoherent part. The
coherent part $G_{coh}$ contains a re-normalization factor given
by the doping level, and does not satisfy the sum rules alone. The
condition on $\phi$ and $\del$ in our case pushes the $p$-fermion
to high energies above the Fermi level, and manifests as the
higher binding energy excitations in the electronic ARPES as we
demonstrated in Figs.1.

In the mean field decoupling of the $J$-term in the hamiltonian,
we replace the projection operators $E^p$ by $(1-x)$ for simplest
purposes.\cite{wen} The remaining decoupling of the $f$-spin
exchange is standard, and we arrive at the following mean field
hamiltonian, assuming translation and spin-rotational invariance,
\eqa H_{MF}=& &\sum_k f_k^\dagger
(\epsilon^f_k\sigma_3+\Del_k\sigma_1)f_k+\sum_k
p_k^\dagger\epsilon^p_k\sigma_3 p_k\nonumber\\ & &
+\sum_k[p_k^\dagger (\xi_k\sigma_3+\eta_k\sigma_1) f_k +{\rm
h.c.}],\eea where we defined the spinors $p_k=(p_{k\ua},
p_{-k\da}^\dagger)^T$ and $f_k=(f_{k\ua}, f_{-k\da}^\dagger)^T$.
Here $ \epsilon^f_k=-(3/4)(1-x)^2\tilde{J}\sum_{n=x,y}\chi_n\cos
k_n+(\phi^2-\del^2)\sum_n \tilde{t}_n\cos k_n\nonumber-\mu_f$ is
the $f$-fermion dispersion, $\Del_k=\sum_n 2\phi\delta
\tilde{t}_n\cos
k_n\nonumber-(3/4)(1-x)^2\tilde{J}\sum_{n=\hat{x},\hat{y}}\Del_n\cos
k_n$ is the $f$-fermion pairing function,
$\epsilon^p_k=\sum_{n\neq \hat{x},\hat{y}} \tilde{t}_n\cos
k_n-\mu_p$ is the $p$-fermion dispersion, $\xi_k=\sum_n \phi
\tilde{t}_n\cos k_n-\lambda\phi$, and finally $\eta_k=\sum_n
\delta \tilde{t}_n\cos k_n-\lambda\del$. In the above expressions
$k_n=\v k\cdot \v r_n$ with $\v r_n$ the vectors along bond $n$,
$\chi_n$ and $\Del_n$ is the $f$-fermion hopping and pairing
amplitudes on bond $n$. Finally $\mu_f$, $\mu_p$ and $\lambda$ are
Lagrangian multipliers that enforce the $f$- and $p$-fermion
occupation and Eq.(\ref{constrain}), respectively. Because of the
constrain on $f$-fermion occupancy, the effective parameters
$\tilde{t}_n$ and $\tilde{J}$ are re-normalized counterparts of
their bare values, $\tilde{t}_n=2t_n$ and $\tilde{J}=4J$, in
similar spirit to RMFT based on Gutzwiller projection of one-band
fermions.\cite{rmft} Note that although the renormalized hopping
integral is twice of the bare value the Mott physics is built in
the quasi-particle-like $d$-operators as we discussed above.
Following ref.\cite{wen}, in the $p$-fermion dispersion we exclude
the nearest neighbor hopping terms in the mean field theory. The
argument is as follows. The bare $p$-fermions are high energy
degrees of freedom, and is thus sensitive to local correlations.
They would view the $f$-spin background as Neel ordered states,
and therefore can only move coherently on the same sublattice.
This is called coherent path approximation.\cite{wen}

We find that the choice of bare parameters $t_{x,y}=0.4eV$,
$t_{x+y,x-y}=0eV$, $t_{2x,2y}=0.06eV$, and $J=0.13eV$ nicely
reproduce the experimental features. The general conclusion is
however not sensitive to parameter tuning. Our self-consistent
calculations yield that $\phi=\del=\sqrt{x/2}$ and $\mu_f=0$, and
that $\chi_n$ changes sign while $\Del_n$ does not in going from
$x$-bond to $y$-bond. These combine to still form a $d$-wave
superconducting pairing
$\Del_{ij}^{sc}=\<d_{i\da}d_{j\ua}-d_{i\ua}d_{j\da}\>_0$. We
stress however that the waterfall feature is indifferent to this
order parameter. For example, the waterfall persists in Fig.1(c)
where $\Del^{sc}\ra 0$. The Matsubara Green's function
$G_d^{\sigma\sigma}(\v k,i\w_n)$ for the $d$-operators, which in
our case is independent of $\sigma$, can be easily obtained from
the mean field hamiltonian. The electronic ARPES spectral function
is then obtained as $A(\v k,\w)=-(2/\pi)\Im G_d^{\ua\ua}(\v
k,i\w_n\ra -\w+i0^+)$,\cite{note} where the minus sign before $\w$
reflects the fact that we worked in the hole-picture in the mean
field theory. The results are shown in Figs.1 and 2, and have been
discussed previously. More details and results will be presented
elsewhere.

To conclude, we proposed a theory that successfully explains the
waterfall feature in the quasi-particle dispersion observed in
hole-doped cuprates. We interpret the waterfall as a connection
between the fermi level and the parent LHB. The theory projects
electrons by quantum interference between two bands of fermions.
The local sum rules satisfied by such a representation make it
possible to pick up the higher binding energy features that is
beyond the scope of mean field theories based on one-band fermions
(but of course not beyond the exact theory of the one-band t-J
model if any).

\acknowledgments{We thank X. G. Wen, X. J. Zhou, Z. Y. Weng and F.
C. Zhang for discussions. QHW thanks the Center of Theoretical and
Computational Physics, the University of Hong Kong for hospitality
in the initial stage of this work. The work was supported by NSFC
10325416, the Fok Ying Tung Education Foundation No.91009, and the
Ministry of Science and Technology of China (973 project No:
2006CB601002).}


\begin{references}
\bibitem{note} A constant elastic scattering rate $\tau^{-1}=0.05eV$ is used in the
calculation of $A(\v k,\w)$. A further phenomenological inelastic
scattering rate could be used to improve the agreement with data
but is not the concern of a mean field theory.
\bibitem{su2} This fact is also reflected in the invariance of the
hamiltonian under an arbitrary local isospin $SU(2)$
transformation $ f_\sigma\ra \alpha f_\sigma+\beta
\sum_{\sigma'}\epsilon_{\sigma\sigma'} f^\dagger_{\sigma'}$
subject to $\sum_\sigma f_\sigma^\dagger f_\sigma=1$ at each site.
Here $|\alpha|^2+|\beta|^2=1$. This symmetry signifies that
$f$-fermions are charge neutral spinons.
\end{references}
\end{document}